# Orbital-resolved anisotropic electron pockets in electron-doped $SrTiO_3$ observed by ARPES

Yuki K. Wakabayashi,[1,a)] Akihira Munakata,[2] Yoshitaka Taniyasu,[1] and Masaki Kobayashi[2,3,a),b)]

[1]NTT Basic Research Laboratories, NTT Corporation, Atsugi, Kanagawa 243-0198, Japan
[2]Department of Electrical Engineering and Information Systems, The University of Tokyo, Bunkyo, Tokyo 113-8656, Japan
[3]Center for Spintronics Research Network, The University of Tokyo, 7-3-1 Hongo, Bunkyo-ku, Tokyo 113-8656, Japan

[a)]**Author to whom correspondence should be addressed:** yuuki.wakabayashi@ntt.com and masaki.kobayashi.brl@ntt.com
[b)]**Present address:** NTT Basic Research Laboratories, NTT Corporation, Atsugi, Kanagawa 243-0198, Japan

## ABSTRACT

$SrTiO_3$ has attracted considerable interest as a wide-band gap semiconductor for advanced high-*k* capacitors and photocatalytic applications. Although previous angle-resolved photoemission spectroscopy (ARPES) studies have characterized the valence band structure originating from O 2*p* orbitals, the conduction band arising from Ti 3*d* orbitals upon electron doping, which is called electron pockets, remain poorly understood. In this study, polarization-dependent ARPES measurements were performed on Nb 1%-doped $SrTiO_3$ (001), enabling direct, orbital-selective visualization of the electron pockets. From the measured band dispersion, we quantitatively determined their effective masses, anisotropy, and electron density. Our results revealed formation of an electron pocket at the Γ point induced by Nb doping, yielding a direct bandgap of 3.79 eV at Γ, consistent with previous optical measurements. Furthermore, the effective masses of $m_1 = 0.63m_0$ (short-axis direction) and $m_2 = 8.0m_0$ (long-axis direction) were identified, where $m_0$ is the free electron mass, and the Fermi surface has been shown to be ellipsoidal. The electron density derived from these dispersions was found to be $3.58 \times 10^{20}$ cm$^{-3}$. These findings provide a comprehensive picture of the conduction-band electronic structure that will be crucial in the design of STO-based functional devices.

$SrTiO_3$ (STO) is an extensively studied perovskite oxide semiconductor because of its large bandgap energy, high dielectric constant, and high chemical stability, making it promising for high-$k$ dielectric applications and photocatalytic applications.[1,2,3,4,5,6,7,8,9] It is also widely used as a substrate for epitaxial growth of various perovskite oxides and serves as a fundamental material in oxide electronics.[10,11,12,13] STO is a perovskite $d_0$-band insulator with a filled O 2$p$ valence band and empty Ti 3$d$ $t_{2g}$ conduction band (CB).[14] The electronic structure of STO, particularly the valence band consisting primarily of O 2$p$ orbitals, has been well-characterized through angle-resolved photoemission spectroscopy (ARPES).[15,16,17] However, the CB states remain less well understood.[18] Upon electron doping, electrons populate the Ti 3$d$ $t_{2g}$ orbitals,[16,19] forming CB electron pockets (EPs) at the Γ point of the Brillouin zone. A detailed understanding of these EPs is essential, as they govern key transport properties such as electron mobility, effective mass anisotropy, and carrier density in STO-based devices. Despite their importance, direct ARPES measurements of these EPs has been lacking.

In this work, we used polarization-dependent ARPES to investigate the CB electronic structure of Nb 1%-doped STO (Nb:STO) single crystal, in which Nb doping introduces electrons into the STO. We directly observed and quantitatively analyzed the orbital-selective EPs, elucidating their effective mass anisotropy, electron density, and polarization characteristics. Our results indicate the EPs have an ellipsoidal Fermi surface with strong mass anisotropy and a direct bandgap of 3.79 eV at the Γ point. These findings clarify previously unresolved aspects of STO's CB structure, providing key insights necessary for tailoring the electronic properties of STO-based electronic and photocatalytic devices.

The polarization-dependent ARPES measurements were performed at BL28 in the Photon Factory, KEK.[20] Circularly polarized ($C$-polarized) and linearly [e.g., horizontally ($E_P$) and vertically ($E_S$)] polarized vacuum ultraviolet (VUV) ($h\nu$ = 50–110 eV) light were used. The experimental geometry is shown in Fig. 1(a). All of the ARPES measurements were performed at 12 K. The total energy resolution at $h\nu$ = 82 eV was of ~27 meV. A clean (001) surface of Nb:STO single crystal (Crystec GmbH) was prepared by *in situ* cleaving under an ultra-high vacuum of $2 \times 10^{-10}$ Torr. The Fermi level ($E_F$) was determined by measuring gold spectra.

The linear polarization dependence of the photoemission process allows for identification of orbital symmetry through the dipole selection rules. Under normal emission geometry [Fig. 1(a)], the photoemission final state $|f\rangle$ possesses even symmetry with respect to the mirror plane of the crystal surface. Consequently, the dipole transition matrix element $\langle f\,|\boldsymbol{A}\cdot\boldsymbol{p}|\,i\rangle$ becomes nonzero only when the initial state $|i\rangle$ has the same parity (even or odd) as the dipole operator $\boldsymbol{A}\cdot\boldsymbol{p}$, where $\boldsymbol{A}$ and $\boldsymbol{p}$ denote the electromagnetic vector potential and the momentum operator, respectively.[21,22,23] For instance, when $P$- ($S$-) polarized light is used, corresponding to an even (odd) parity of $\boldsymbol{A}\cdot\boldsymbol{p}$, only the initial states with matching symmetry contribute to the ARPES signal. This polarization-dependent selection rule enables the orbital characters to be disentangled, as is schematically summarized for Ti 3$d$ $t_{2g}$ orbitals in Fig. 1(b).

As STO exhibits band dispersion along $k_z$ due to its three-dimensional electronic structure, photon energy-dependent ARPES measurements are essential in order to identify the location of the Γ point in momentum space. Figure 2(a) illustrates the Brillouin zone of STO, where the red plane indicates the Γ–X–M–X plane corresponding to the $k_z$-Fermi surface (FS) mapping in Fig. 2(b). The $k_z$ FS map in Fig. 2(b) obtained with $C$-polarized light clearly shows a three-dimensional EP centered at the Γ point, consistent with the Nb-induced electron doping of the CB, confirming that the observed states originate from bulk CBs rather than surface two-dimensional states.[24,25,26,27] On the basis of the $k_z$ dispersion, we identified a photon energy of 82 eV to correspond to the Γ–X plane and used this energy in the subsequent measurements.

Figure 3(a) shows the ARPES spectra of Nb:STO obtained with *C*-polarized light along the Γ–X direction. In addition to the O 2*p* valence bands located between approximately 4 and 8 eV below $E_F$,[16] the intensity in the vicinity of $E_F$ at the Γ point arises from the Ti 3*d* $t_{2g}$ states, which is occupied by dopant electrons. In the energy distribution curve (EDC) at the Γ point [Fig. 3(b)], a dispersive quasiparticle peak is observed at $E_F$. In the prototypical metallic perovskite oxide $SrVO_3$, oxygen vacancies are known to give rise to a non-dispersive incoherent peak about 1.5 eV below $E_F$;[28] the absence of any such feature here confirms that we are probing the intrinsic electronic structure with negligible influence from oxygen vacancies. The energy position of the valence band maximum is estimated to be 3.85 eV [Left inset in Fig. 3(b)] below $E_F$. The energy difference between this valence band maximum and the conduction band minimum at 60 meV below $E_F$, as estimated from the EP band dispersion (discussed later), yields a Γ-Γ direct band gap of 3.79 eV, which is in good agreement with the value of 3.75 eV estimated from optical measurements for STO.[29,30] It should be noted that STO is an indirect-gap semiconductor with a Γ–M indirect band gap of 3.3 eV.[15] Previous ARPES studies on STO reported a low-binding-energy shoulder structure at the O 2*p* valence band maximum and attributed this feature to in-gap states caused by localized carriers near oxygen vacancies[17] or strong electron-phonon coupling,[16] both of which hinder accurate determination of the band gap from ARPES measurements. In contrast, such a shoulder structure is absent in the present study, likely due to our use of a high-quality cleaved surface of a single crystal. Indeed, optical measurements have shown that in-gap absorption occurs in STO thin films when defects are intentionally introduced.[7]

To investigate the detailed electronic structure of the EP at the Γ point, polarization-dependent ARPES measurements were performed. Figures 4 shows the high-resolution ARPES spectra along the Γ–X direction measured with *C*-, *S*-, and *P*-polarized light. Under *C*-polarized light [Fig. 4(a)], two distinct bands appear: one with a light effective mass and the other with a heavy effective mass. On the basis of the orbital selection rules for linearly polarized light as illustrated in Fig. 1(b), the heavy mass band is attributed to the $d_{yz}$ orbital [Fig. 4(b)], while the light mass band corresponds to the $d_{xz}$ orbital [Fig. 4(c)]. The contribution from the $d_{xy}$ orbital is not clear from the *S*-polarization measurements, which is consistent with photoemission matrix element calculations that take the emission angle into account,[31] indicating that in the present experimental geometry, the photoemission intensity associated with the $d_{xy}$ orbital is weaker than that of the $d_{yz}$ orbital due to its predominantly in-plane orbital character. In order to see the band dispersions more clearly, we took the second derivatives of the ARPES spectra, as shown in the lower panels of Figs. 4(b) and 4(c). The effective masses of these bands were determined by fitting quadratic curves to the second derivatives. We also estimated the Fermi energy relative to the conduction band minimum, $\varepsilon_F$, to be 60 meV, from the bottom of the quadratic fitting curves. The light mass band exhibited an effective mass of $m_1 = 0.63m_0$, while the heavy mass band showed an effective mass of $m_2 = 8.0m_0$, where $m_0$ denotes the free electron mass. These experimentally determined values differ by nearly a factor of two from those predicted by density functional theory calculations ($m_1 = 1.1m_0$ and $m_2 = 4.4m_0$),[32] highlighting the difficulty of making quantitative theoretical predictions on strongly correlated systems and the importance of experimentally determining the properties of EPs. These effective masses, which have been quantitatively determined from direct observations of the anisotropic electron pockets, should facilitate more precise heterojunction designs and contribute to the development of STO-based functional devices.

The mass anisotropy of EPs has not been quantitatively determined by making electrical measurements such as Hall or thermoelectric measurements, because the observed values correspond to weighted averages of $m_1$ and $m_2$. The experimentally determined $m_1$ and $m_2$ values in this study will serve as a basis for quantitative theoretical analysis of previously reported anisotropic electronic properties, including charge transport [33] and microwave dielectric response.[34] Furthermore, in the two-dimensional electron gas where the Ti 3*d* $t_{2g}$ orbitals are quantized along the *z*-direction, the effective masses $m_1$ = 0.67–0.7 and $m_2$ = 9.7–20 have been estimated from the band dispersion measured by ARPES.[25,27] These values are slightly larger than those of the 3D bulk EPs obtained in this study, suggesting the presence of a mass renormalization due to electronic correlations.[35,36]

Under the effective-mass approximation, the anisotropic band dispersion of the $t_{2g}$ electron pocket can be expressed as

$$E(k_x, k_y, k_z) = \frac{\hbar^2}{2m_1}({k_x}^2 + {k_y}^2) + \frac{\hbar^2}{2m_2}{k_z}^2, \tag{1}$$

where $\hbar$ is the reduced Planck constant, and $k_x$, $k_y$, and $k_z$ are the wavevectors along the *x*, *y*, and *z* directions, respectively. Considering the sixfold degeneracy due to the spin and orbital degrees of freedom for the $t_{2g}$ bands, the electron carrier density *n* of Nb:STO can be estimated by integrating over the occupied state as follows:

$$n = \frac{1}{(2\pi)^3}\iiint_{E<\varepsilon_\mathrm{F}} E(k_x, k_y, k_z)\, d\boldsymbol{k} = \frac{2^{\frac{3}{2}}}{18\pi^2}\frac{\sqrt{{m_1}^2 m_2}}{\hbar^3}\varepsilon_\mathrm{F}. \tag{2}$$

Inserting $\varepsilon_\mathrm{F}$ = 60 meV, $m_1 = 0.63m_0$, and $m_2 = 8.0m_0$ into Eq. (2) yields $n = 3.6\times10^{20}$ cm$^{-3}$. This value is approximately twice the nominal Nb donor density ($1.7\times10^{20}$ cm$^{-3}$), assuming one electron is donated per Nb atom substituting for Ti in the lattice.[37,38] Nevertheless, the two values are of the same order of magnitude, suggesting that the ARPES-based estimate is reasonably accurate. The modest deviation may originate from several factors, including additional electron donation due to defects introduced during cleaving, and the band bending effect near the surface.[18]

Figure 5 shows the FSs of the Ti 3*d* $t_{2g}$ EPs, simulated using the above-determined values of $m_1$, $m_2$, and $\varepsilon_\mathrm{F}$, alongside experimentally observed FSs under different polarization conditions. The polarization-dependent measurements enable orbital-selective mapping of the FS, revealing distinct anisotropic shapes associated with the $d_{xz}$ and $d_{yz}$ orbitals, while the $d_{xy}$ contribution remains suppressed due to its in-plane orbital character and corresponding matrix element effects, as described earlier.[31] On the basis of these direct observations of band dispersions and FSs, we quantitatively determined the effective masses, FS geometry, and carrier density of the EPs in *n*-type STO. These results constitute the first comprehensive orbital-resolved determination of these characteristics in electron-doped STO by using ARPES.

In summary, we performed polarization-dependent ARPES measurements on Nb:STO to directly probe the orbital-resolved electronic structure of the CBs. Our analysis revealed anisotropic EPs centered at the Γ point, with distinct contributions from the Ti 3*d* $t_{2g}$ orbitals. From the band dispersions, we quantitatively extracted the effective masses, FS geometry, and carrier density, which was found to be consistent within an order of magnitude of the nominal Nb doping level. These results will provide a critical foundation for understanding and engineering electronic transport in STO-based functional devices.

**ACKNOWLEDGMENT**

This work was supported by Grants-in-Aid for Scientific Research (24H00018) from the Japan Science and Technology Agency. This work was partially supported by the Spintronics Research Network of Japan (Spin-RNJ). The measurements at KEK-PF were performed under the approval of the Program Advisory Committee (Proposal No. 2024G558 and 2024S2-001) at the Institute of Materials Structure Science at KEK.

## CONFLICT OF INTEREST

The authors have no conflicts of interest to disclose.

## AUTHORS' CONTRIBUTIONS

**Y. K. Wakabayashi**: Conceptualization (equal); Validation (lead); Investigation (lead); Supervision (equal); Writing – Original Draft (lead); Writing – Review & editing (equal). **A. Munakata**: Investigation (supporting); Writing – Review & editing (supporting). **Y. Taniyasu**: Writing – Review & editing (supporting). **M. Kobayashi**: Conceptualization (equal); Validation (supporting); Investigation (supporting); Supervision (equal); Writing – Review & editing (equal).

## DATA AVAILABILITY

The data that support the findings of this study are available from the corresponding author upon reasonable request.

## Figures and Tables

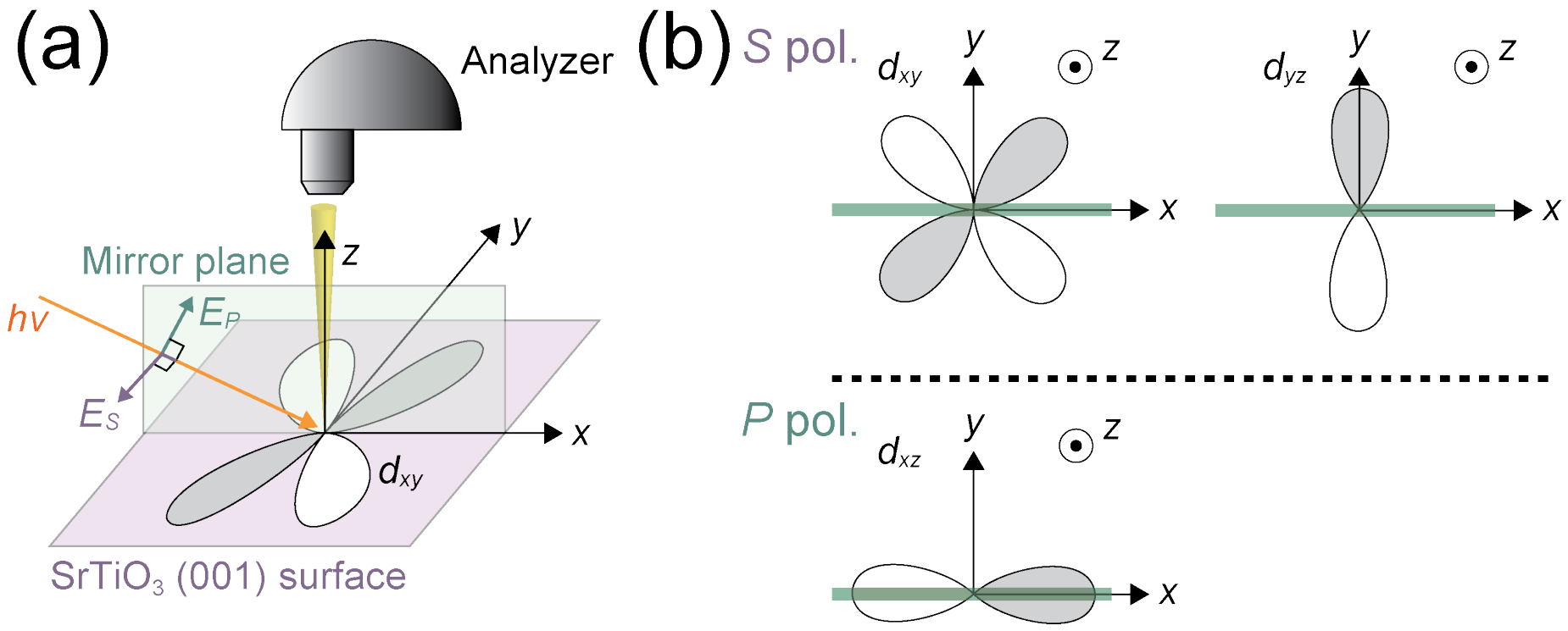

**Fig. 1.** (a) Schematic diagram of the experimental setup of the polarization-dependent ARPES measurements. (b) Spatial symmetry of the $3d$ $t_{2g}$ orbitals with respect to the mirror plane, which are selectively excited with $S$ and $P$ polarizations. The mirror plane including the direction of incidence and emission is defined along the $x$-axis, as shown by the green lines.

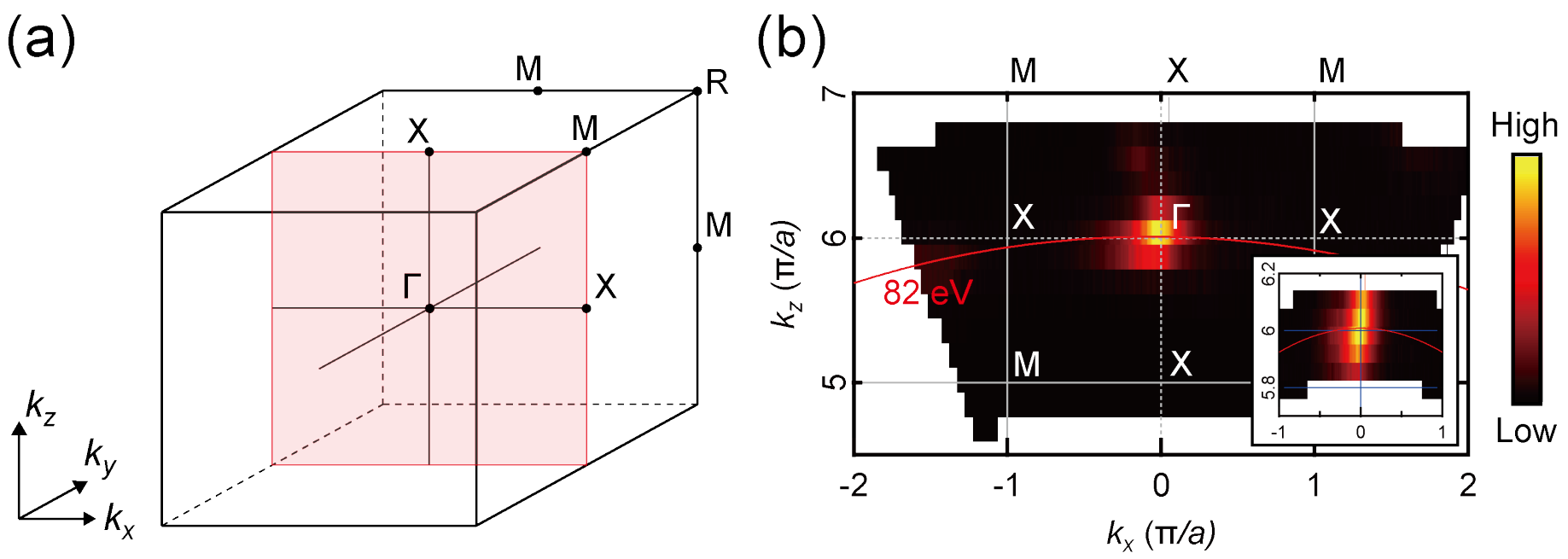

**Fig. 2.** (a) Brillouin zone of cubic $SrTiO_3$. (b) Fermi surface of Γ-X-M-X plane, as indicated in (a) for Nb:$SrTiO_3$, obtained from the $C$ polarization measurements and integrating over a binding energy window of +20 to −20 meV. The photon energy varies from 50 to 110 eV in 4 eV steps. The inset shows a more detailed Fermi surface near the Γ point, measured in 1 eV steps from 77 eV to 87 eV. The red lines indicate the trace in momentum space of $h\nu$ = 82 eV. $k_z$ is calculated with an inner potential $V_0$ of 11 eV and STO lattice constant of 3.905Å.

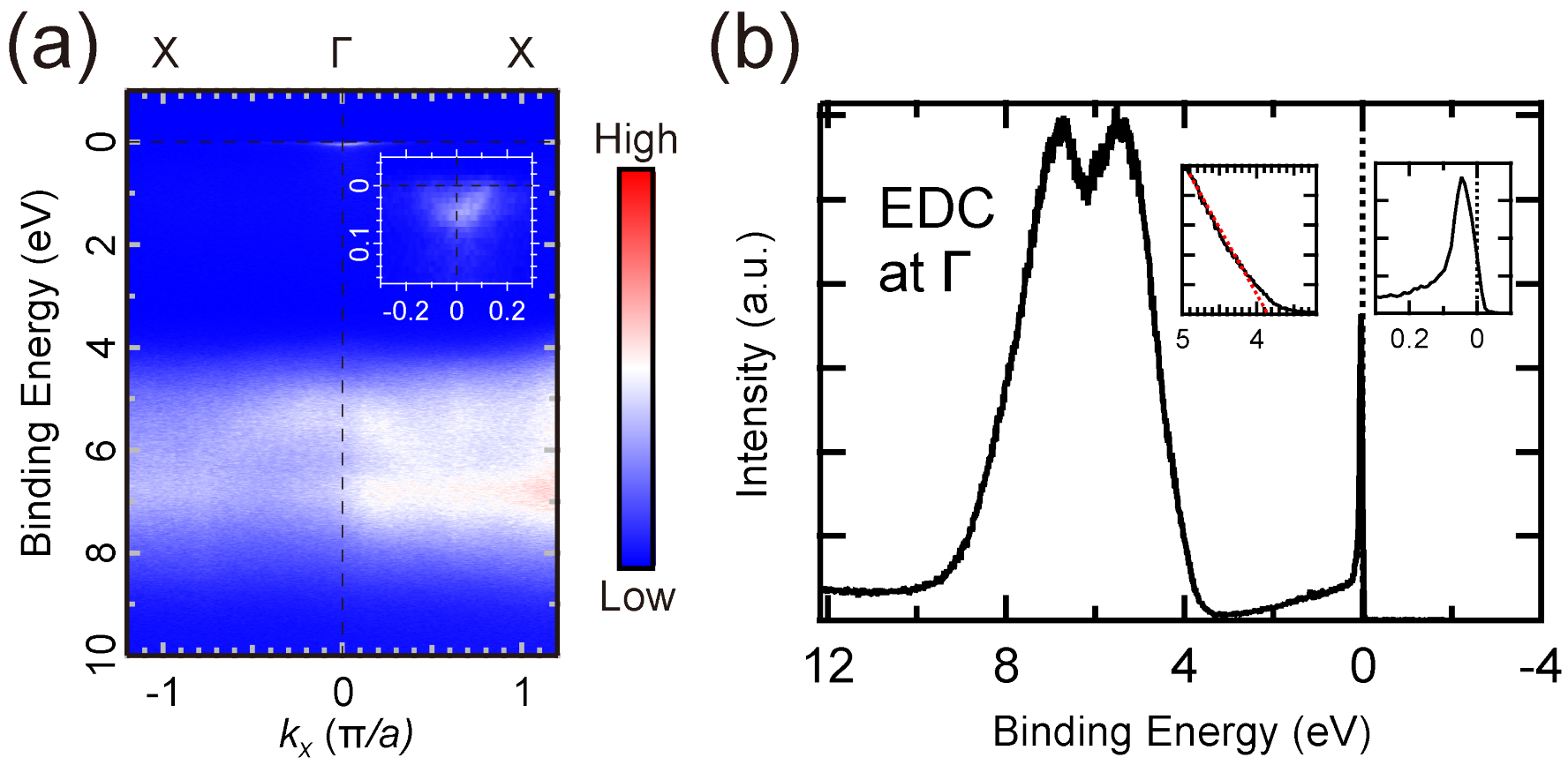

**Fig. 3.** (a) ARPES spectra along the Γ-X direction for Nb:$SrTiO_3$ obtained from *C* polarization measurements. The inset in (a) is a magnified image at the Γ point. (b) Photoemission spectrum at Γ [($k_x$, $k_y$, $k_z$) = (0,0,0)] for Nb:$SrTiO_3$ obtained from the *C* polarization measurements. In (b), the left inset shows a linear extrapolation of the photoemission intensity (red dashed line) onset at the Γ point. It was used to estimate the energy of the valence band maximum as defined by the intersection with the zero-intensity baseline. In (b), the right inset is a magnification of the spectrum at $E_F$.

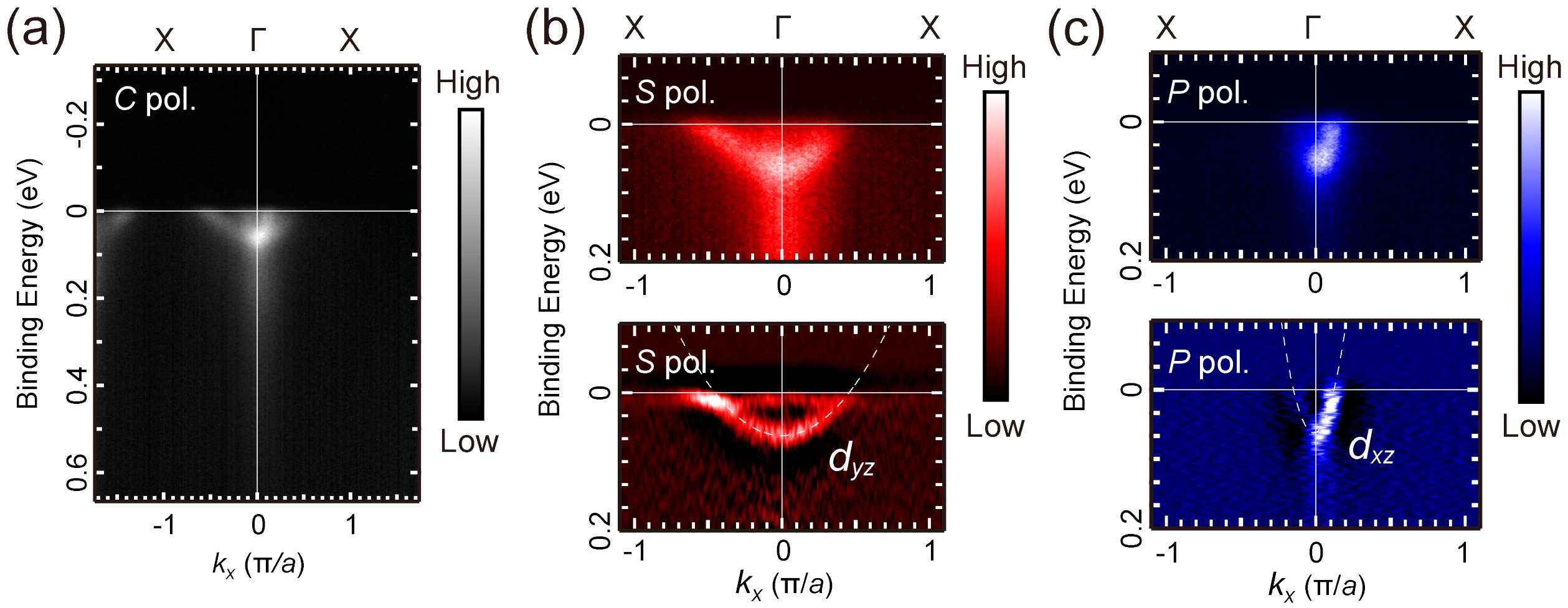

**Fig. 4.** ARPES spectra at the Γ point for Nb:$SrTiO_3$ obtained from (a) *C*, (b) *S*, and (c) *P* polarization measurements. The lower panels in (b) and (c) are the corresponding second derivatives. The dashed lines in the lower panels of (b) and (c) indicate quadratic fitting curves.

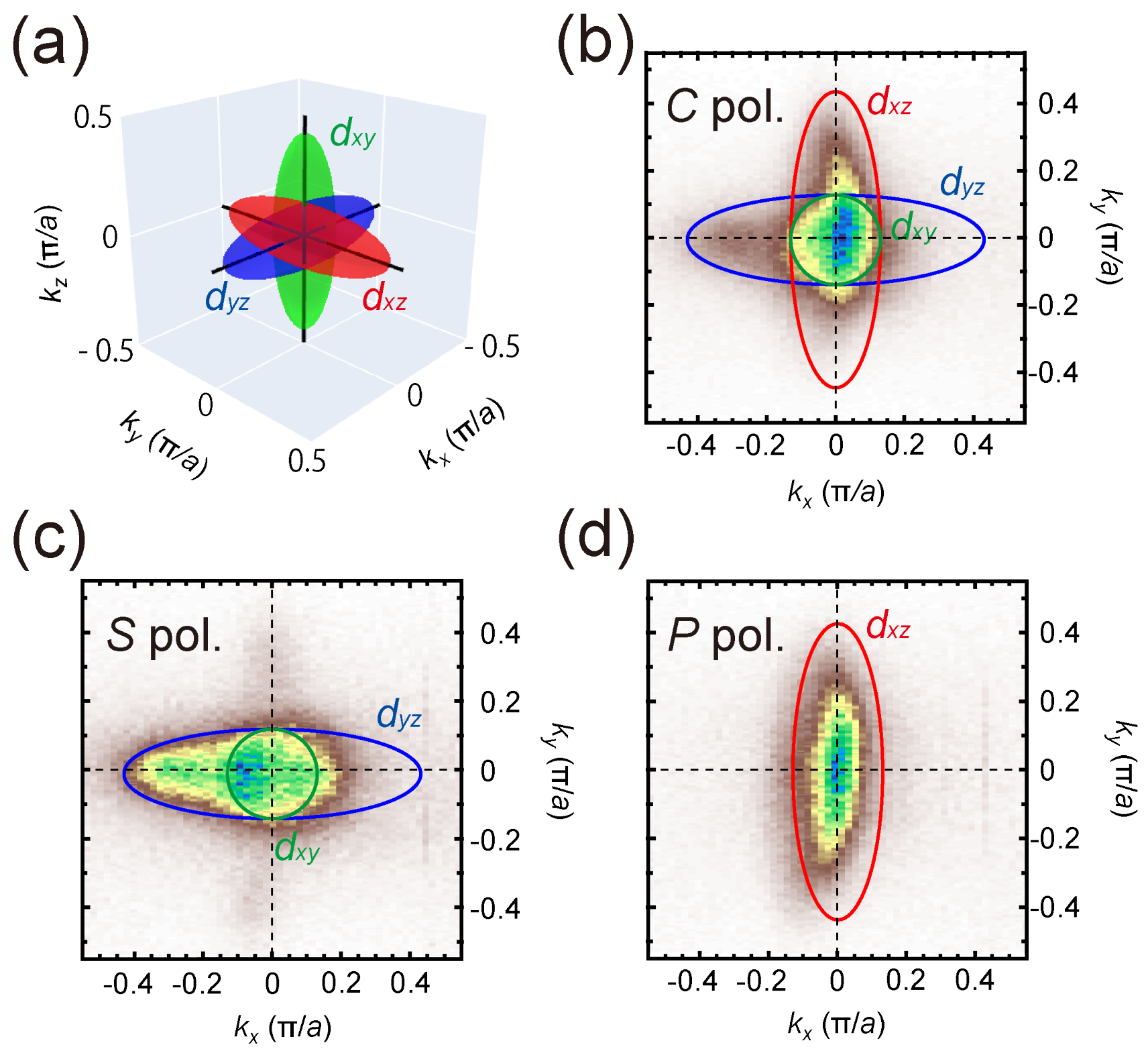

**Fig. 5.** Simulated Fermi surfaces of the Ti $t_{2g}$ orbitals having $m_1 = 0.63m_0$ and $m_2 = 8.0m_0$. (b)-(d) Fermi surfaces in the $k_x$-$k_y$ plane including the Γ point for Nb:$SrTiO_3$ obtained from (b) *C*, (c) *S*, and (d) *P* polarization measurements. In (b)-(d), red, blue, and green ellipses and circles represent the Fermi surfaces of the $d_{xz}$, $d_{yz}$, and $d_{xy}$ orbitals, respectively.

.